\documentclass[aps, reprint, amsmath, amssymb, superscriptaddress, longbibliography, pra]{revtex4-2}
\usepackage{graphicx} 
\usepackage{physics}
\usepackage{tabularx}
\usepackage{upgreek}

\usepackage{layouts} 
    
\usepackage{amsmath}
\usepackage{hyperref}
\hypersetup{
 colorlinks=true,
 linkcolor=blue,
 anchorcolor = blue,
 citecolor = blue,
 filecolor = blue,
 urlcolor = blue
}

\usepackage{bbm}
\usepackage{dcolumn}
\usepackage{braket}
\usepackage{bm}
\usepackage{color}
\usepackage{verbatim}
\usepackage{comment}
\usepackage{dsfont}
\usepackage{physics}
\usepackage{float}
\usepackage{ stmaryrd } 
\usepackage{siunitx} 
\usepackage[percent]{overpic} 
\renewcommand{\selectlanguage}[1]{} 

\newcommand{\logical}{\mathrm{L}}
\newcommand{\physical}{\mathrm{phy}}

\newcommand{\qubit}{\ensuremath{\mathrm{QB}}}
\newcommand{\data}{\ensuremath{\mathrm{D}}}
\newcommand{\ancilla}{\ensuremath{\mathrm{A}}}

\newcommand{\ztype}{\ensuremath{\mathrm{Z}}}
\newcommand{\xtype}{\ensuremath{\mathrm{X}}}
\newcommand{\fourtwotwo}[0]{\ensuremath{\left\llbracket 4,2,2 \right\rrbracket}}
\renewcommand{\bell}{\ensuremath{\Phi}}

\newcommand{\move}[0]{\ensuremath{\mathrm{MOVE}}}
\newcommand{\iswap}[0]{\ensuremath{\mathrm{iSWAP}}}
\newcommand{\cz}[0]{\ensuremath{\mathrm{CZ}}}
\newcommand{\fidReadout}[0]{\ensuremath{F_\mathrm{RO}}}
\newcommand{\fidSQBGsim}[0]{\ensuremath{F_\mathrm{SQ,sim}}}
\newcommand{\fidSQBGind}[0]{\ensuremath{F_\mathrm{SQ,ind}}}
\newcommand{\fidMOVE}[0]{\ensuremath{F_\mathrm{m}}}
\newcommand{\fidCZ}[0]{\ensuremath{F_\mathrm{cz}}}
\newcommand{\pSQBG}[0]{\ensuremath{P_\mathrm{SQBG}}}
\newcommand{\pMOVE}[0]{\ensuremath{P_\mathrm{MOVE}}}
\newcommand{\pCZ}[0]{\ensuremath{P_\mathrm{CZ}}}
\newcommand{\pReadout}[0]{\ensuremath{P_\mathrm{RO}}}
\newcommand{\cycletime}[0]{\ensuremath{t_\mathrm{cycle}}}
\newcommand{\ptherm}[0]{\ensuremath{P_\mathrm{q}}}
\newcommand{\Ttherm}[0]{\ensuremath{T_\mathrm{q}}}
\newcommand{\ler}{\ensuremath{\varepsilon}}
\newcommand{\DHex}{six-qubit star lattice}

\begin{document}
\title{Quantum error detection in qubit-resonator star architecture}
\author{Florian Vigneau}
\author{Sourav Majumder}
\author{Aniket Rath}
\affiliation{IQM Quantum Computers, Georg-Brauchle-Ring 23-25, 80992 Munich, Germany}
\author{Pedro Parrado-Rodr\'iguez}\affiliation{IQM Quantum Computers, P. de la Castellana 200, Madrid 28046, Spain}
\author{Francisco Revson Fernandes Pereira}
\author{Hsiang-Sheng Ku}
\author{Fedor Šimkovic IV}
\author{Stefan Pogorzalek}
\author{Tyler Jones}
\author{Nicola Wurz}
\author{Michael Renger}
\author{Jeroen Verjauw}
\author{Ping Yang}
\author{William Kindel} 
\author{Frank Deppe}
\affiliation{IQM Quantum Computers, Georg-Brauchle-Ring 23-25, 80992 Munich, Germany}
\author{Johannes Heinsoo}
\affiliation{IQM Quantum Computers, Keilaranta 19, 02150 Espoo, Finland}
\date{\today}

\begin{abstract}

Achieving industrial quantum advantage is unlikely without the use of quantum error correction (QEC).
Other QEC codes beyond surface code are being experimentally studied, such as color codes and quantum Low-Density Parity Check (qLDPC) codes, that could benefit from new quantum processing unit (QPU) architectures.
We introduce the \DHex{} architecture that offers parallelism and effective local all-to-all connectivity and thus enables hardware-efficient implementation of certain QEC codes.
As a first demonstration of this new architecture, we encode two logical qubits in a six-qubit superconducting QPU with a star-topology using the \fourtwotwo{} code and characterize the logical states with the classical shadow framework. Logical life-time and logical error rate are measured over repeated quantum error detection cycles for various logical states including a logical Bell state.
We measure logical state fidelities above \SI{96}{\percent} for every cardinal logical state, find logical life-times above the best physical element, and logical error-per-cycle values ranging from \SIrange{0.25 \pm 0.02}{0.91 \pm 0.03}{\percent}. 
In future, such star QPU can be tiled to enable QEC codes with high-weight and overlapping stabilizers for improved encoding rates.

\end{abstract}

\maketitle	

\section{Introduction}

There are industrially relevant computational tasks which can be solved faster, cheaper, or more accurately using quantum computers, but achieving industrial quantum advantage likely requires error rates only accessible by means of quantum error correction (QEC)~\cite{beverland_assessing_2022,preskill_beyond_2025}.
In stabilizer based QEC~\cite{Gottesman1997,Terhal2015}, two-qubit gates are used to map data-qubit parity to ancilla qubit states~\cite{fowler2012surface}.
Repeated measurements of the ancilla qubit states provide error syndrome data, which can be decoded to identify likely errors and extend the lifetime of the logical qubit.
Logical error rates depend on the physical quantum gate fidelity, the physical qubit lifetime, the topological properties of the used code, and also on the match between code structure and architecture of the QPU.
While the planar square grid of superconducting qubits fits well the stabilizers of surface code~\cite{Kitaev1998,fowler2012surface,Wallraff2020,Google2024}, there are
other stabilizer circuits and codes such as planar color codes~\cite{bombin2006topological} and qLDPC codes~\cite{IBM2024_qLDPC,eberhardt2024pruningqldpccodesbivariate}, which could outperform surface code on other QPU architectures in terms of logical error rate for a given number of physical qubits~\cite{gidney2023superdense}. 

Compared to surface codes, large-distance color codes and qLDPC codes have additional requirements~\cite{Rice2011,Fujii2024,IBM2024_qLDPC}. First, they require the measurement of at least weight-six stabilizers. Secondly, X and Z basis stabilizers are supported by the same data qubits. Thus, color codes do not map to a locally connected 2D array of qubits in a trivial way. This difficulty has been overcome in quantum computers based on trapped ions~\cite{Monz2022, Monz2024}, where two-qubit gates are possible between qubits that share a vibrational mode of the trap, and on neutral atoms~\cite{Grangier2007,Lukin2024}, where atoms can be shuttled around modifying the effective connectivity during operation. With superconducting qubits, individual stabilizers of color code have been demonstrated on a five qubit device \cite{Takita2017,Vuillot2018,Flammia2019} and a mapping of color code has been proposed for the heavy-hex architecture~\cite{Chamberland2020} where extra two-qubit gates and qubits were employed for routing and measuring syndrome-circuit error flags. Recently, a distance-three color code has been demonstrated on a square grid architecture~\cite{Google2024ColorCode} using a superdense coding circuit~\cite{gidney2023superdense}. To execute qLDPC codes, long-range couplers that cross over other elements or make use of additional routing surfaces have been proposed~\cite{MIT2025qLDPC} and demonstrated~\cite{,wang2025qLDPC}.

The smallest possible color code is the \fourtwotwo{}~code~\cite{Vaidman_1996,Grassl_1997,Yamamoto1997,Gottesmann2016}, which protects two logical qubits in four data qubits. The code allows correction of erasure errors and detection of up to one error on the data qubits. Many experimental demonstrations~\cite{Vuillot2018, Michielsen2018, Flammia2019, Monroe2017, Hanzo2021, Tucker2024} followed the encoding proposal of Ref.~\cite{Gottesmann2016} using a 5-qubit ring, while others used stabilizer measurements~\cite{Takita2017}. The \fourtwotwo{}~code has been used to demonstrate variational quantum eigensolver algorithms with increased accuracy~\cite{Jong2020,OakRidge2024} and magic state injection~\cite{Gupta2024}. The code allows logical single- and two-qubit gates within the same patch with simple physical level operations~\cite{Shapiro2018}. A CNOT gate between the logical qubits in the same code can be done by swapping logical qubits, which, at least in absence of other code-patches, can also be done virtually by relabeling qubits. A two-qubit gate in between codes can be done by transversal CNOT gates~\cite{AtomComputing2024} or logical parity measurements according to Ref.~\cite{landahl2014quantum}. Recently, using the same code, a 24-logical-qubit cat state was prepared using 12 pairs of logical qubits encoded in the state of 256 neutral atoms~\cite{AtomComputing2024}.

To enable qubit- and gate-count-efficient encoding with color and qLDPC codes on superconducting qubit platforms, increased connectivity needs to be provided in a scalable fashion. 
Star lattices, constructed from cells with star connectivity, address this need within a single 2D plane. They combine high local connectivity stemming from their star character with the scalability and parallelism inherent to their lattice character. Depending on the chosen design details, such as the number of qubits per cell, these two characteristics can be traded off against each other to best suit the envisioned application. 
Generally, the cells can be made from any type of single-star topology implementations~\cite{Vijay2021Ring,Heimonen2022,Cleland2024Modular,Orcutt2024Bell,Deneb,Fan2025Mpemba} where a group of qubits is connected via a central element or a shared mode enabling low-circuit-depth two-qubit interactions for one pair at a time.

\begin{figure}[h!]
    \begin{overpic}{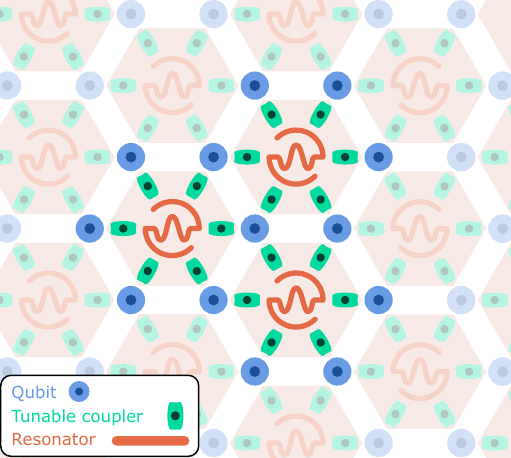}    
    \end{overpic}     
    \caption{Illustration of the \DHex{} architecture with a resonator as central element. Apart from the boundary region, every central resonator connects to six qubits via respective tunable couplers and every qubit is coupled to three resonators. Through these resonators, each qubit is able to interact with its 12 nearest-neighbor qubits as shown in the highlighted sublattice.}
    \label{fig:Star_Lattice}
\end{figure}

As a specific example, we propose the \DHex{} architecture shown in Fig.~\ref{fig:Star_Lattice}, which is constructed from hexagonal cells with qubits in the corners. These qubits are coupled by tunable couplers to a central element. 
Then, every qubit inside the lattice connects to its 12 nearest neighbors via one of the three coupled central elements. Although the central elements can only be used for one qubit-qubit interaction at the time, the ratio of qubits and central elements is two to one. In other words, all qubits can simultaneously participate in one of 12 groups of two-qubit gates (see more details in App.~\ref{Supp_section:Star_lattice}). Thus, on star lattice QPUs, it is possible to efficiently measure spatially overlapping and high-weight stabilizers as required by color and qLDPC codes. Furthermore, the \DHex{} architecture can be advantageous in the context of NISQ applications such as quantum simulations ~\cite{papivc2025near,algaba2025fermion,leppakangas2025quantum}.

To demonstrate the quantum error correction potential of the \DHex{} architecture, we perform an error detection experiment with a device featuring a single star. We demonstrate logical error detection by encoding two logical qubits with the \fourtwotwo{} code in a six-qubit-star QPU. 
To start, we describe the error detection scheme in Sec.~\ref{sec: code and circuit}, characterize the stabilizer code in Sec~.~\ref{sec:stabilizer tomo}, and use the classical shadow technique~\cite{Huang2020,elben_review} to measure the quality of the logical state encoding in Sec.~\ref{sec:tomography}. We then repeat the error detection cycle to measure the logical state decay in Sec.~\ref{sec:logical error rate}. To take it further, we measure the decay of a logical two-qubit Bell state in Sec.~\ref{sec:bell}. We finally conclude on the performance of the error detection scheme in Sec.~\ref{sec:conclusion}.

\section{Description of the error detection scheme}\label{sec: code and circuit}

\begin{figure*}
    \begin{overpic}[scale=1.0]{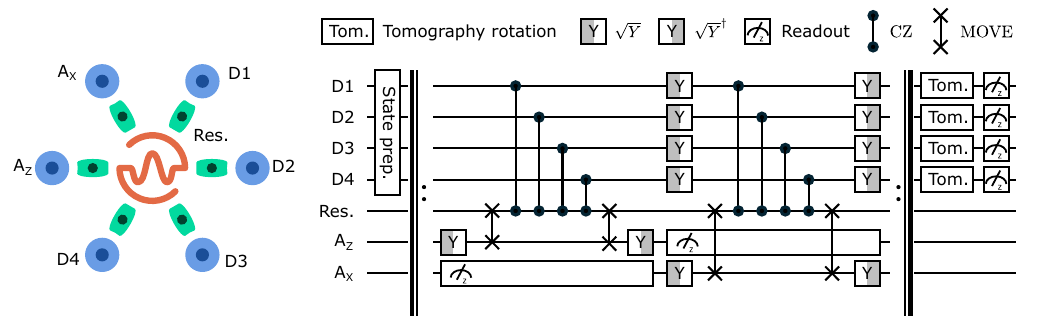}    
        \put(0,28){(a)}    
        \put(27,28){(b)}
    \end{overpic}     
    \caption{a) Schematic of the six-qubit-star QPU featuring four data qubits (\data{}1--\data{}4) and two ancilla qubits ($\mathrm{\ancilla{}_\xtype{}}$ and $\mathrm{\ancilla{}_\ztype{}}$) connected via tunable couplers (green) to a central resonator (Res., orange).
    b) Gate sequence of the distance-2 error detection cycle consisting of state preparation sub-circuit; stabilizer measurement cycles implemented with single qubit $\sqrt{Y}$ gates, its inverse $\sqrt{Y}^\dagger$, resonator-qubit $\cz{}$ and $\move{}$ gates and ancilla qubit measurements; single qubit gates for state tomography and final data qubit readout operations, see App.~\ref{Supp_section:Detailed_QEC_circuit}. stabilizer measurement cycles can be repeated $N$ time, with the $\ancilla{}_\xtype{}$ measurement of cycle $n$ starting at the beginning of cycle $n+1$.}
    \label{fig:Sample_QEC_Cyle}
\end{figure*}

The QPU used in this work~\cite{Deneb}, illustrated in Fig.~\ref{fig:Sample_QEC_Cyle}a, features six transmon qubits each connected by a tunable coupler to the central resonator. Our coupler and gate schemes allow us to work around common challenges such as frequency crowding, frequency targeting, and static ZZ coupling~\cite{Oliver2018Coupler,Marxer2023Coupler}.
We use four data qubits $\data{}1-\data{}4$ to encode two logical states and the two remaining qubits $\ancilla{}_\xtype{}$ and $\ancilla{}_\ztype{}$ as ancillae for error syndrome detection.

The code is defined by the stabilizers
\begin{align}
\begin{split}
\label{eq:stabilizers}
    S_\xtype{} &= X_{\data{}1}X_{\data{}2}X_{\data{}3}X_{\data{}4}, \\
    S_\ztype{} &= Z_{\data{}1}Z_{\data{}2}Z_{\data{}3}Z_{\data{}4},
\end{split}
\end{align}
where $X_i$ and $Z_i$ are Pauli operators for qubit $i$, and by the logical operators of the two logical qubits
\begin{equation}    
\begin{array}{r@{}l}
\label{eq:log_operators}
    &X_{\logical{}1} = X_{\data 1}X_{\data 3} \text{   ,   } Z_{\logical{}1} = Z_{\data 1}Z_{\data 2}, \\
    &X_{\logical{}2} = X_{\data 1}X_{\data 2} \text{   ,   } Z_{\logical{}2} = Z_{\data 1}Z_{\data 3},
\end{array}
\end{equation}
which in turn define the logical codewords that relate the physical states to the logical states, see App.~\ref{Supp_section:CodeWord}.

The physical gate circuits studied in this paper consist of state preparation, stabilizer measurement cycle and data qubits readout, see Fig.~\ref{fig:Sample_QEC_Cyle}b. State preparation circuit varies for experiments described in this paper. We use stabilizer measurements to encode separable logical states, we prepare the data qubits in an initial state $\psi_{\mathrm{in}}$ that overlaps with a single codeword $\psi_{\mathrm{target}}$ using natural qubit decay and single-qubit gates before the first cycle. Then, in the absence of errors, the first measurement of the stabilizers projects the data qubits to the corresponding codeword state with $1/2$ probability (probabilistic encoding strategy). State preparation for the logical Bell state is described in Sec.~\ref{sec:bell}.

Our stabilizer measurement circuit consists of \num{12} single-qubit $\sqrt{Y}$ gates, ten qubit-resonator gates, and two ancilla qubit readout operations, see Fig.~\ref{fig:Sample_QEC_Cyle}b. First, $\sqrt{Y}$ gates are used to change between Z and X basis. Then a \move{} operation is used to transfer the state from an ancilla qubit to the central resonator, where \move{} is similar to an \iswap{} gate, but limited to the state-space spanned by $\ket{0g},\ket{1g},\ket{0e}$ states of the resonator-qubit subsystem~\cite{Deneb}. Next, $\cz{}$ gates are successively applied between data qubits and the central resonator. Finally, the state is moved back to the ancilla qubit with a second \move{}. The ancilla qubit is then measured to obtain the value of the stabilizer. A similar gate sequence is realized with the second ancilla to measure the other stabilizer. This second sequence can be executed while the first ancilla qubit is being measured, following an interleaved measurement scheme inspired by Ref.~\cite{Versluis2017}. 
After $N$ error detection cycles, data qubits are measured. In the tomography experiments, combinations of single qubit gates from the set $\{I,\sqrt{X},\sqrt{Y}\}$ are applied before data qubit readout.

This circuit for stabilizer measurement is not fault tolerant in a sense that a single bitflip error in the computational resonator in the middle of the syndrome extraction circuit can lead to an undetectable logical error. Adding an additional qubit as a flag would allow detecting these errors~\cite{chao2018flag}, 
see App.~\ref{Supp_section:Flag}. However, for higher distance color codes, there are syndrome extraction circuits without error flags which outperform a circuit with imperfect-flags as well as surface code for the same resources and error model~\cite{gidney2023superdense}. Thus, it is still an open question which circuit and architecture is optimal for future QEC codes.

\section{Stabilizer characterization}
\label{sec:stabilizer tomo}

\begin{figure}[h!]
    \begin{overpic}[scale=1.0]{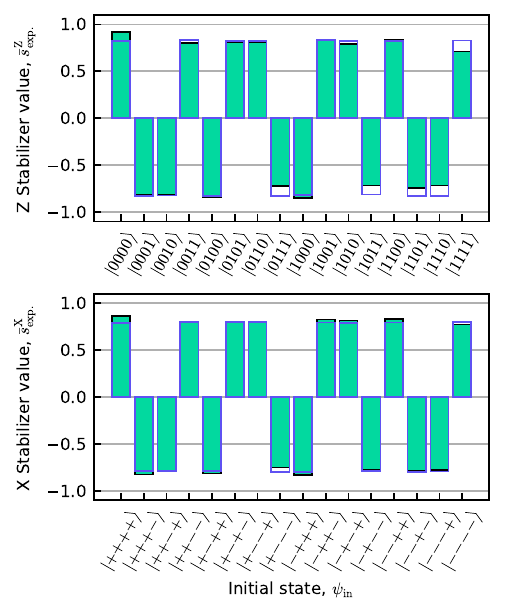}    
        \put(-1,97){(a)}    
        \put(-1,52){(b)}
    \end{overpic} 
    \caption{Single stabilizer measurement expectation values $\bar{s}_{\rm exp.}^\ztype{}$ (a) and $\bar{s}_{\rm exp.}^\xtype{}$ (b) for initial state $\psi_{\mathrm{in}}$, averaged over $10^5$ repetitions from experiment (green) and error-model simulation (blue). $\ket{\pm} = \left(\ket{0}\pm\ket{1}
\right)/\sqrt{2}$.}
    \label{fig:Stabilizer_Characterization}
\end{figure}

We characterize the performance of the stabilizers measurement circuits by measuring the expectation values of the individual stabilizers in a single stabilizer tomography experiment~\cite{Wallraff2022}. 
We study a single stabilizer at a time by executing only first or second half of the circuit of the cycle given in Fig.~\ref{fig:Sample_QEC_Cyle}b. 
We repeat the experiment with \num{16} separable states $\psi_{\mathrm{in}}$ which form an eigenbasis of the studied stabilizer such that we expect to measure $s_{i,\rm ideal}=\bra{\psi_{i,\rm in}}S\ket{\psi_{i,\rm in}}=\pm 1$. The observed stabilizer values averaged over experiment repetitions $\bar{s}_{\rm exp.}$, see Fig.~\ref{fig:Stabilizer_Characterization}, are well reproduced by the individually characterized errors, detailed in App.~\ref{Supp_section:simulation}. We observe a stabilizer fidelity~\cite{Wallraff2022} $1-\langle \lvert \bar{s}_{i,\rm exp.} - s_{i,\rm ideal} \rvert \rangle /2=\SI{90.1}{\percent}$ averaged over the input states $\psi_{i,\mathrm{in}}$ for $S_\xtype{}$ and \SI{89.8}{\percent} for $S_\ztype{}$. 

\section{Quantum state tomography} \label{sec:tomography}

We perform tomography experiments to understand the performance of the logical and physical state preparation. 
To derive unbiased estimations of quantities of interest in this work, we employ the framework of classical shadows, also known as shadow tomography, based on local Pauli measurements~\cite{Huang2020,elben_review}, see App.~\ref{Supp_section:RM} for further information.

In this experiment, we prepare $\psi_{\rm in}$ from the set of \num{16} separable states, for which either $\bra{\psi_{\rm in}}S_{\xtype}\ket{\psi_{\rm in}}=1$ or $\bra{\psi_{\rm in}}S_{\ztype}\ket{\psi_{\rm in}}=1$. Then, we measure both stabilizers once using the full circuit given in Fig.~\ref{fig:Sample_QEC_Cyle}b. Thereafter, for each initial state, we apply to the data qubits all of the $3^4 = 81$ combinations of single-qubit Pauli tomography gates. 
For each instance of these rotations, we repeat the experiment \num{2e3} times.

The measurement data is post-selected conditioned on the result of both stabilizer measurements $s^\ztype{}$ and $s^\xtype{}$ yielding a value of ${+}1$. Due to our choice of input and target states, we would reject \SI{50}{\percent} of states even in the ideal case. In addition, there are errors in data qubits and stabilizer circuit resulting in \textit{survival rate} $P_\mathrm{S} \in [\num{0.65},\num{0.90}]$, see Table~\ref{tab:logical_metrics}, and total post selection yield is $P_\mathrm{S}/2$.

\begin{table}[b]
\centering
\renewcommand{\arraystretch}{1.2}
\setlength{\tabcolsep}{5pt} 
\begin{tabular}{llccccc}
\hline
\textbf{$\psi_{\mathrm{in}}$} & $\psi_{\mathrm{target}}$ & $\ F_\logical{}$ & $p_{2,\logical{}}$ & $p_{2,\physical{}}$ & $\ P_\logical{}$ & $ P_{\rm S}$\\ 
\hline
$\ket{0000}$ & $\ket{00}_\logical{}$ & 0.989 & 0.876 & 0.322 & 0.723 & 0.848 \\
$\ket{1111}$ & $\ket{00}_\logical{}$ & 0.992 & 0.880 & 0.299 & 0.714 & 0.832 \\
$\ket{0011}$ & $\ket{01}_\logical{}$ & 0.978 & 0.855 & 0.294 & 0.714 & 0.874 \\
$\ket{1100}$ & $\ket{01}_\logical{}$ & 0.993 & 0.879 & 0.295 & 0.713 & 0.852\\
$\ket{0101}$ & $\ket{10}_\logical{}$ & 0.979 & 0.853 & 0.293 & 0.730 & 0.526 \\
$\ket{1010}$ & $\ket{10}_\logical{}$ & 0.975 & 0.848 & 0.289 & 0.721 & 0.862 \\
$\ket{0110}$ & $\ket{11}_\logical{}$ & 0.999 & 0.918 & 0.285 & 0.606 & 0.852 \\
$\ket{1001}$ & $\ket{11}_\logical{}$ & 0.982 & 0.872 & 0.289 & 0.667 & 0.878 \\
 $\ket{{-}{-}{-}{-}}$ & $\ket{{+}{+}}_\logical{}$ & 0.978 & 0.854 & 0.268 & 0.677 & 0.848\\
 $\ket{{+}{+}{+}{+}}$ & $\ket{{+}{+}}_\logical{}$ & 0.968 & 0.837 & 0.274 & 0.694 & 0.902\\
 $\ket{{+}{-}{+}{-}}$ & $\ket{{+}{-}}_\logical{}$ & 0.966 & 0.831 & 0.274 & 0.709 & 0.648\\
 $\ket{{-}{+}{-}{+}}$ & $\ket{{+}{-}}_\logical{}$ & 0.981 & 0.857 & 0.278 & 0.698 & 0.870\\
 $\ket{{-}{-}{+}{+}}$ & $\ket{{-}{+}}_\logical{}$ & 0.993 & 0.882 & 0.276 & 0.681 & 0.676\\
 $\ket{{+}{+}{-}{-}}$ & $\ket{{-}{+}}_\logical{}$ & 0.981 & 0.861 & 0.274 & 0.682 & 0.870\\
 $\ket{{+}{-}{-}{+}}$ & $\ket{{-}{-}}_\logical{}$ & 0.966 & 0.834 & 0.274 & 0.693 & 0.846\\
 $\ket{{-}{+}{+}{-}}$ & $\ket{{-}{-}}_\logical{}$ & 0.968 & 0.836 & 0.277 & 0.702 & 0.904\\
 \hline
\end{tabular}
\caption{Summary of results extracted from shadow tomography for different initial physical states $\psi_{\mathrm{in}}$ that encode logical states $\psi_{\mathrm{target}}$ in X and Z basis respectively. 
}
\label{tab:logical_metrics}
\end{table}

We construct the density matrix of the logical state $\rho_\logical{} = \sum_{i, j} \bra{i} \rho_\physical \ket{j}/ P_\logical{}$ by projecting the physical state $\rho_\physical$ onto the logical subspace $\ket{i}$, $\ket{j}$ $\in \{\ket{00}_\logical{},\ket{01}_\logical{},\ket{10}_\logical{},\ket{11}_\logical{} \}$ with logical \textit{acceptance probability} $ P_\logical{} = \bra{i} \rho_\physical \ket{i}$ \cite{Wallraff2020}. 
Finite probability $P_\logical{} \in [\num{0.606}, \num{0.730}]$ indicates that there are errors not captured by syndrome measurements. The probability $P_\logical{}$ also relates logical and physical state fidelity $F_\logical=\bra{\psi_{\mathrm{target}}}\rho_{\logical}\ket{\psi_{\mathrm{target}}}=F_\physical/P_\logical{}$~\cite{Gilchrist2005,Wallraff2020}. We observe logical fidelity $F_\logical{} > \num{0.965}$ for all the prepared two-qubit logical states. 
Slightly lower fidelities for X basis states arise from the additional single qubit gates required to prepare $\psi_{\mathrm{in}}$. The total success probability of producing and measuring a state in the logical subspace is $\eta = (1/2)P_\mathrm{S} P_\logical{}$.

Purity of the logical state $p_{2,\logical{}} = \Tr(\rho_\logical^2) \in [\num{0.83}, \num{0.92}]$ indicates finite probability of always preparing the same logical state. As physical state purity $p_{2,\physical{}} \in [\num{0.27}, \num{0.32}]$, is considerably lower than $p_{2,\logical{}}$ for all input states, we have confirmed that projection to the logical subspace rejects states prepared out of logical subspace.

The reconstructed physical and logical density matrices obtained by quantum state tomography are shown in App.~\ref{Supp_section:logical_tomo}.

\section{Repeated error detection} \label{sec:logical error rate}

\begin{figure*}
    \includegraphics[scale=1.0]{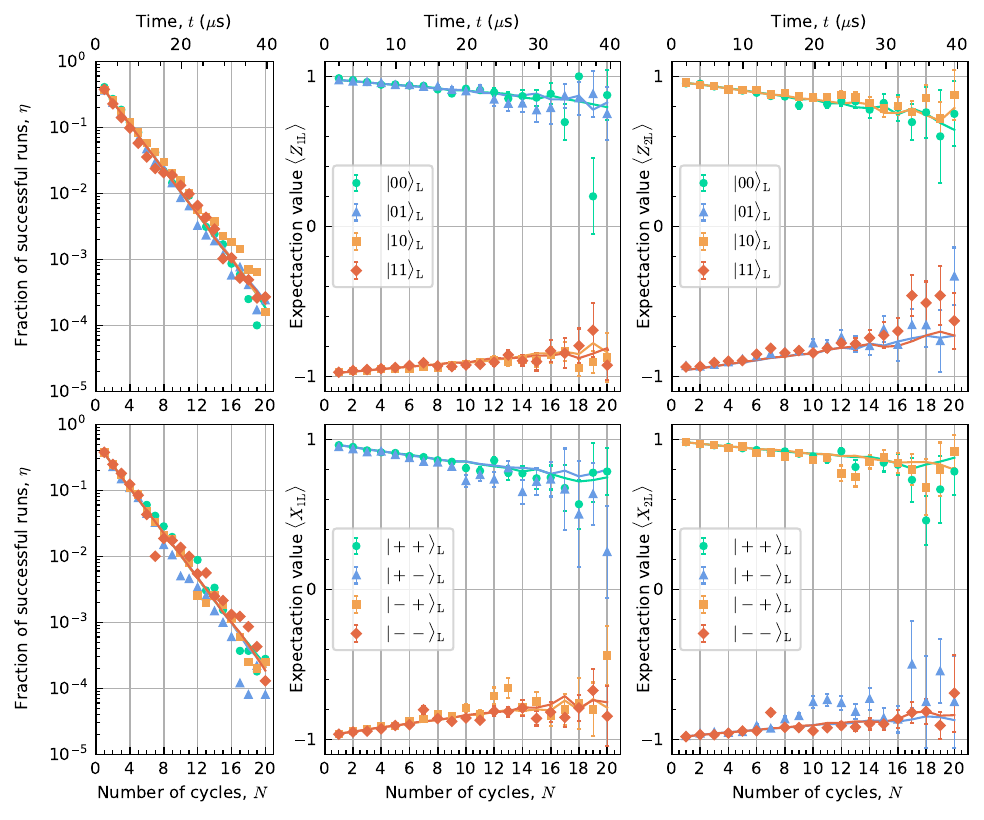}
    \put(-470,380){(a)}    
    \put(-470,190){(b)}
    \put(-335,380){(c)}    
    \put(-335,190){(d)}
    \put(-170,380){(e)}    
    \put(-170,190){(f)}
    \caption{
    a,b) Fraction of successful run $\eta$ where no error have been detected versus number of cycle $N$ and time $t$, for the logical states in caption.
    c-f) Expectation values of the logical operators $Z_{\logical{}1}$, $Z_{\logical{}2}$, $X_{\logical{}1}$ and $Z_{\logical{}2}$ versus number of cycle. The error bars are one standard deviation calculated considering a binomial distribution with $\eta_N$. 
    Solid lines are simulations.
    \label{fig:Logical_expectation_values}}
\end{figure*}

Next, to obtain the logical lifetime and error rate, we conduct lifetime experiments with four target X and Z basis states, by preparing $\ket{0000}$, $\ket{0011}$, $\ket{0101}$, $\ket{1001}$, $\ket{{+}{+}{+}{+}}$, $\ket{{+}{-}{+}{-}}$, $\ket{{+}{+}{-}{-}}$ and $\ket{{-}{+}{+}{-}}$ with $\ket{\pm} = \left(\ket{0}\pm\ket{1}
\right)/\sqrt{2}$.
We repeat the error detection cycle for $N\in [1,20]$. 
For each error detection cycle $n \leq N$, we measure both stabilizers $s^\xtype_n$ and $s^\ztype_n$. 
After $N$ cycles, we measure the data qubit states in the same basis as the preparation. 
We post-select the measurement results based on the stabilizers not detecting any errors and data qubits being assigned to the logical subspace before evaluating $\langle X_{\logical{}i} \rangle_N$ and $\langle Z_{\logical{}i} \rangle_N$ of logical qubit $i$ from $10^5$ experiment repetitions. For comparison, we simulate $10^6$ repetitions of the experiment with Stim~\cite{gidney2021stim}, see App.~\ref{Supp_section:simulation} for details.

We find that the number of accepted runs follows an expected relation $\eta_N = P_{\mathrm{S}}^N P_{\logical{}}^{\phantom{N}}/2$ and agrees with the simulation, see Fig.~\ref{fig:Logical_expectation_values}a and Fig.~\ref{fig:Logical_expectation_values}b. We estimate $P_\mathrm{S}=\eta_N/\eta_{N-1} \approx\num{0.67}$ and $P_\logical{}=2\eta_1/P_\mathrm{S}\approx \num{0.87}$. These values are consistent with those derived from the tomography experiments in the previous section. More information on syndrome detection probability is available in App.~\ref{Supp_section:detectors} and on leakage dynamics in App.~\ref{Supp_section:Leakage}.

Also, the decay dynamics of the expectation values of $Z_{\logical{}1}$, $Z_{\logical{}2}$, $X_{\logical{}1}$ and $X_{\logical{}2}$ agrees with the simulation. We fit a function $a e^{-N \cdot b}$, where $a$ and $b$ are free parameters and use the successful run fraction $\eta_N$ to weight the data points, see Figs.~\ref{fig:Logical_expectation_values}c-f. We calculate logical lifetime $\tau_{\ztype} = \cycletime/b \in \left[138,407\right]~\si{\micro\second}$ from the experiment in Z basis and logical coherence time $\tau_{\xtype} \in \left[111,249\right]~\si{\micro\second}$ for X basis, see Table \ref{table:Logical_life_time}. The observed logical life- and coherence times are all improved compared to the best physical component, see App.~\ref{Supp_section:qubit_characteristics}. 

From the fitted parameters, we also obtain the error rate, logical error per cycle $\ler = (1-e^{-b})/2$, which corresponds to a decay of the expectation value 
$\braket{X_{\logical{}i}}_{N+1}= ( 1-2\ler_{Xi})\braket{X_{\logical{}i}}_{N}$ and
similar for $Z_\logical{}$~\cite{Google2023}. The error rate ranges from $\ler_{\ztype{}1}=\SI{0.25}{\percent}$ for $\ket{10}_\logical$ to $\ler_{\ztype{}1}=\SI{0.91}{\percent}$ for $\ket{{-}{+}}_\logical$, see Table~\ref{table:Logical_life_time}. All logical error rates are lower than error rate $1-\fidCZ{} \approx \SI{1.03}{\percent}$ of the best CZ gate used in the experiment.

\begin{table}[thb]
\centering
\renewcommand{\arraystretch}{1.2}
\setlength{\tabcolsep}{5pt} 
\begin{tabular}{cccccc}
\hline
\textbf{State}           &   $\tau_{\ztype{}1}$ (µs)     &   $\tau_{\ztype{}2}$  (µs)  & $\ler_{\ztype{}1}$ (\%) &   $\ler_{\ztype{}2}$ (\%) \\
\hline
$\ket{00}_\logical{}$    &   \SI{175 \pm  9}{}   &   \SI{157 \pm 1}{}  &  \SI{0.58 \pm 0.03}{}    &  \SI{0.64 \pm 0.05}{}  \\
$\ket{01}_\logical{}$    &   \SI{303 \pm 15}{}    &   \SI{204 \pm 38}{}   &  \SI{0.34 \pm 0.02}{}    &  \SI{0.5 \pm 0.08}{}   \\
$\ket{10}_\logical{}$    &   \SI{407 \pm 34}{}    &   \SI{155 \pm  8}{}   &  \SI{0.25 \pm 0.02}{}    &  \SI{0.65 \pm 0.03}{}  \\
$\ket{11}_\logical{}$    &   \SI{242 \pm 15}{}    &   \SI{138 \pm 12}{}   &  \SI{0.42 \pm 0.02}{}    &  \SI{0.73 \pm 0.06}{}  \\
    \\
\hline
    & $\tau_{\xtype{}1}$ (µs)      & $\tau_{\xtype{}2}$ (µs)     & $\ler_{\xtype{}1}$ (\%) &   $\ler_{\xtype{}2}$ (\%) \\
\hline
$\ket{{+}{+}}_\logical{}$&   \SI{129  \pm 5}{}     &   \SI{190  \pm  7}{}   &  \SI{0.79  \pm 0.03}{}    &  \SI{0.53  \pm 0.02}{}  \\
$\ket{{+}{-}}_\logical{}$&   \SI{134  \pm 9}{}     &   \SI{192  \pm 24}{}   &  \SI{0.76  \pm 0.04}{}    &  \SI{0.53  \pm 0.06}{}  \\
$\ket{{-}{+}}_\logical{}$&   \SI{111  \pm 4}{}     &   \SI{178  \pm 14}{}   &  \SI{0.91  \pm 0.03}{}    &  \SI{0.57  \pm 0.04}{}  \\
$\ket{{-}{-}}_\logical{}$&   \SI{153  \pm 8}{}     &   \SI{249  \pm 25}{}   &  \SI{0.66  \pm 0.03}{}    &  \SI{0.41  \pm 0.04}{}  \\
\hline
\end{tabular}
\caption{\label{table:Logical_life_time} Logical lifetime $\tau_{\ztype{}i}$, coherence time $\tau_{\xtype{}i}$, logical $X_\logical{}$ and $Z_\logical$ error probability $\ler_{\xtype{}i}$ and $\ler_{\ztype{}i}$ for logical qubit $i$ for various logical states together with the uncertainty obtained from fit residuals. 
}
\end{table}

These logical error rates are affected by errors in the resonator that can cause undetected correlated errors on data qubits. In App.~\ref{Supp_section:Flag} and~\ref{Supp_section:Comparison} we show how adding flag qubits can detect these errors and improve the performance. We also extrapolate the performances for higher stabilizer weights and compare them with superdense coding.

\section{Logical Bell states}\label{sec:bell}

Here, we demonstrate preparation and preservation of entanglement for two logical qubits.
Between the previous experiment and the following one, the device was thermally cycled and we exchanged the role of some of the physical qubits, see~App.~\ref{Supp_section:qubit_characteristics}.

\newcommand{\ketbell}{\ensuremath{\ket{\bell{}}_{\logical{}}}}
We create a logical Bell state
\begin{equation}
    \ketbell = \frac{\ket{00}_{\rm \data{}1\, \data{}4} + \ket{11}_{\rm \data{}1 \,\data{}4}}{\sqrt 2} \otimes \frac{\ket{00}_{\rm \data{}2 \,\data{}3} + \ket{11}_{\rm \data{}2 \,\data{}3}}{\sqrt 2}
\end{equation}
by creating physical Bell states between the data qubits pairs \data{}1--\data{}4 and \data{}2--\data{}3~\cite{Gottesmann2016}. We prepare the physical Bell pairs with effective \cz{} gate between the qubit pair and $\sqrt{Y}$ gates. Unlike previously, the $\ketbell$ state is encoded by the state preparation circuit already before the first cycle of error detection. Therefore, the experiment does not suffer from rejection of half of the runs by the first stabilizer measurement. Together with the first stabilizer measurement, this state preparation scheme is as fault tolerant as the stabilizer circuit~\cite{Gottesmann2016}.

We repeat error detection cycles with postselection according to the stabilizer measurements and the final physical state being in the logical subspace. To address the decay of the number of post-selected results while also saving time on the experiment with small number of cycles $N$, we scale the number of experiment repetition according to $\num{5e3}N$ and $\num{5e4}N$ for simulation. 

\begin{figure}
    \includegraphics[scale =1.0]{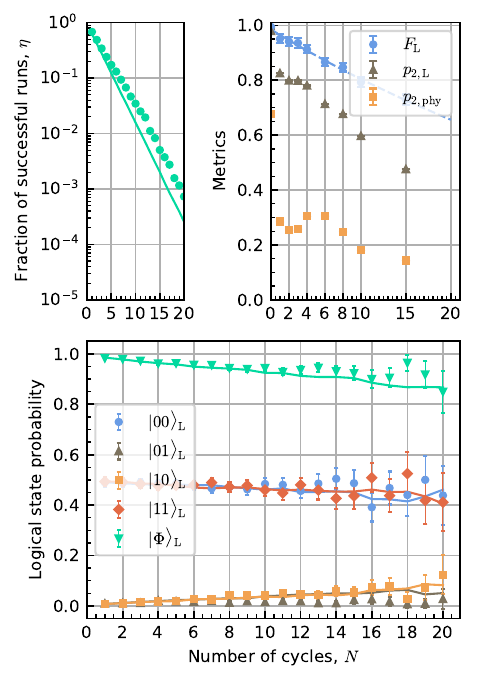}
    \put(-230,320){(a)}    
    \put(-230,160){(b)}
    \put(-130,320){(c)}
    \caption{Repeated error detection for a logical Bell state. 
    (a) Fraction of successful runs during repeated error detection cycles. 
    (b) Logical state probability measured (markers) and simulated (solid lines) over number of cycles for the states $\ket{00}_\logical{}$, $\ket{01}_\logical{}$, $\ket{10}_\logical{}$, $\ket{11}_\logical{}$ and $\ketbell$. The whiskers represent one standard deviation considering a binomial distribution of $\eta_N$. 
    Time is calculated from the duration of the error detection cycle. 
    (c) Logical fidelity $F_\logical{}$, logical purity ${ p_{2, L}}$, physical purity $p_{2,\physical{}}$ of the Bell state given by classical shadow. }
    \label{fig:Bell}
\end{figure}

We see in experiments and simulations qualitatively the same trends of success probability $\eta_N$, see Fig.~\ref{fig:Bell}a. Apparently the success rate in the experiment outperforms our simulations quantitatively, which is likely due to changes in the performance between independent characterization and experiment.
By fitting $\eta_N$ to the experiment data, we find that on average $P_\mathrm{S}=\SI{71}{\percent}$ of the runs are kept after each cycle, which is slightly better than for separable logical states, with an acceptance probability $P_\logical{} \approx 0.88$ close to the observation for other states (see Section~\ref{sec:logical error rate}). 

When measuring the Bell state in Z basis, we observe $\ket{00}_\logical{}$ and $\ket{11}_\logical{}$ states with a probability close to 0.5 each as expected, see Fig.~\ref{fig:Bell}b. We also plot the sum of the observation probabilities of these two states, which an upper bound for probability of logical state $\ketbell$. We observe general quantitative agreement between simulation and experiments.
From an exponential fit to the experimentally obtain Bell state observation probability we obtain the logical life time $\tau_\bell = \SI{400 \pm 30}{\micro\second}$ and logical error per cycle $\ler_\bell = \SI{0.25 \pm 0.02}{\percent}$ which are comparable to the best preserved separable logical state. 

However, Z basis measurement alone is not enough to demonstrate entanglement between the two logical qubits as it cannot distinguish a Bell state from mixed states with classical correlations. To reveal the entanglement between the two logical qubits, we again conduct logical state tomography.

We first characterize the encoding circuit by conducting the tomography right after the initialisation of $\ketbell$ state and before doing any error detection cycles, see App.~\ref{Supp_section:logical_tomo} for the extracted density matrices. The logical fidelity $F_\logical{}=\num{0.995}$, logical purity $p_{2,\logical{}}=\num{0.98}$ and physical purity $p_{2,\physical{}}=\num{0.676}$ are higher than in the previous experiments due to lower number of used physical gates.

We also conduct tomography of the $\ketbell$ state after up to $N=15$ cycles of error detection and post selection.
We scale the number of repetition to $10^3 N$.
We obtain the expected exponential decay of $F_\logical{}$ with logical life time \SI{100 \pm 5}{\micro\second} and logical error rate $\ler_\bell{}=\SI{1.01 \pm 0.04 }{\percent}$, see Fig.~\ref{fig:Bell}c.
Even after $N=\num{15}$ cycles fidelity $F_\logical{} > 0.5$ which indicates that the two logical qubits preserve entanglement. 
We also observe an exponential decay for the logical and physical purity as the state gets more and more mixed, with the exception of $N=0$ discussed before.

\section{Conclusion}\label{sec:conclusion}

We have introduced the general concept of 2D star lattices as a scaling path for single-star topologies in quantum computing. Focusing on the specific example of the \DHex{} architecture for superconducting QPUs, we have demonstrated that the increased local connectivity between the qubits facilitates the execution of complex error-correcting codes. Specifically, we have presented quantum error detection experiments using the \fourtwotwo{} code on a minimal device featuring one central resonator connected to six transmon qubits by tunable couplers.

By encoding a logical Bell state, we have demonstrated entanglement between the two logical qubits. From the state tomography we have revealed logical fidelity from \SIrange{96.6}{99.9}{\percent} for all characterized logical states, which is in line with prior results using the same code on a neutral atom~\cite{Tucker2024} and superconducting qubits~\cite{Takita2017} platform. We have demonstrated preservation of the logical state up to 20 cycles with logical life time $>\SI{100}{\micro\second}$, which is above the life-time of the best physical components of the QPU. The observed logical error rate, characterized by a logical error per cycle ranging from \SIrange{0.25 \pm 0.02}{0.91 \pm 0.03}{\percent} for all cardinal states and \SI{1.01 \pm 0.04 }{\percent} for logical Bell state which is below error probability of the best used physical \cz{} gate. The performance can be improved using flag qubits to detect errors in the resonator, see App.~\ref{Supp_section:Flag} and App.~\ref{Supp_section:Comparison}. Our results exceed those from previous error detection experimental demonstrations using transmon qubits in Ref.~\cite{Takita2017} and Ref.~\cite{Wallraff2020}, but do not reach the state of the art for logical error rate obtained with higher distance surface code~\cite{Google2024} and color code~\cite{Google2024ColorCode}. 

In addition to good coherence of the physical qubits, the performance of the code has been enabled by the high connectivity of the star topology and parallel operation of X and Z ancilla qubits.We have used a resonator as a central element, mapped the data qubit parity to the resonator state with \cz{} gates and swapped ancilla qubit and resonator states using \move{} operations. Our implementation only needs two \move{} operations per ancilla and error detection cycle and which is a modest additional circuit depth. As shown in the error budget in App.~\ref{Supp_section:error_budget}, the \move{} operation contribution to the overall rejection rate and logical error rate is very small. Furthermore, this overhead does not scale with increasing number of data qubits~connected to the same resonator.
In the \DHex{} architecture, any two-qubit gate between nearest neighbor's qubits can be achieve with a sequence of MOVE-CZ-MOVE. In our QPU, the fidelity of the sequence is within $[\num{0.950},\num{0.983}]$ (see App.~\ref{Supp_section:qubit_characteristics}). Although the exact threshold of QEC codes implemented in this architecture is not yet known, more optimization is required to reach performance below the best-known threshold for color codes $p_\mathrm{th} \approx 0.005$ \cite{Google2024ColorCode} and qLDPC codes $p_\mathrm{th} \approx 0.007$ \cite{IBM2024_qLDPC}.

\begin{acknowledgments}
Correspondence regarding this work should be addressed to Florian Vigneau (florian.vigneau@meetiqm.com), Johannes Heinsoo (johannes@meetiqm.com) or Frank Deppe (frank.deppe@meetiqm.com).
We thank Brian Tarasinski, Amin Hosseinkhani, Olexiy Fedorets, Jakub Mrożek and Antti Vepsäläinen for useful discussions.
This work was supported by the German Federal Ministry of Education and Research through the projects DAQC (13N15686), QSolid
(13N16155) and Q-Exa (13N16065). 

\end{acknowledgments}

\bibliography{biblio.bib}

 \onecolumngrid

\newpage
\appendix
\section{Connectivity of \DHex{} architecture}\label{Supp_section:Star_lattice}

\begin{figure*}[h!]
    \centering
    \includegraphics{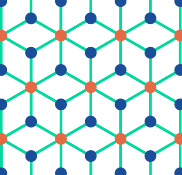}
    \hspace{14pt}
    \includegraphics{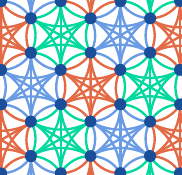}    
    \put(-210,78){(a)}    
    \put(-103,78){(b)}
    \caption{(a) A representation of \DHex{} connectivity as a bipartite rhombille graph formed of qubit- (blue) and central- element- (orange) vertices and coupler edges (green). (b) The same architecture but central elements considered as multiqubit couplers, each corresponding to a hyperedge (15 grouped edges of the same color) illustrating the parallelism of the effective qubit-qubit gates.}
    \label{fig:star_lattice_graph}
\end{figure*}

A formal description of connectivity of the QPU elements is necessary for automated circuit transpilers and performance estimation~\cite{linke_experimental_2017,Holmes2018Impact,nash_quantum_2020,deb_exploring_2021,kotil_improved_2023}. A connectivity graph is defined by a set of edges connecting circuit elements at graph vertices. At a given point in time, a gate is executable if the corresponding vertices and edges are unused.
As shown in Fig.~\ref{fig:star_lattice_graph}a, within the \DHex{} architecture, circuit elements form a \textit{bipartite rhombille} graph with the qubits and the central elements represented by the two sets of vertices (blue and orange respectively), and the couplers represented by edges (green). Within this graph, every qubit can interact with any of its three neighboring central elements via a MOVE or a CZ gate. Multiple CZ gates can be applied between a resonator and different qubits in between two MOVE gates.

The operation sequence MOVE-CZ-MOVE with a single CZ gate effectively achieves a CZ gate between two qubits connected to the same central element. This effective gate allows one to consider the central element including all its tunable couplers as one multi-qubit coupler. Then, the central elements can be eliminated from the connectivity graph and the resulting graph can be considered to be a \textit{six-uniform hypergraph} with hyperedges connecting six vertices each, see Fig.~\ref{fig:star_lattice_graph}b. If one reduces the size of the hyperedges down to three size-two edges, one obtains a sparser graph where faces are hexagons and thus one can also call the full graph a \textit{dense hex}.
The representation of qubit-qubit coupling as hyperedges captures the constraint on the parallelism: at a given time a single qubit-qubit gate can use a single multi-qubit coupler. The hypergraph has 12 perfect matchings, meaning that there are 12 groups of two-qubit gates which can be executed in parallel making use of all qubits which are sufficiently far away from the lattice boundaries. There is such a parallel gate group for each pair of qubits connected via the same multi-qubit coupler.

\section{Codewords of the \fourtwotwo{} code}\label{Supp_section:CodeWord}

Codewords are physical states which correspond to logical states. For \fourtwotwo{} code, for our choice of stabilizers, Eq.~\ref{eq:stabilizers}, and logical operators, Eq.~\ref{eq:log_operators}, codewords for cardinal states of the two logical qubits are

\begin{equation}    
\begin{array}{r@{}l}
\label{eq:codwordsZZ}
    \ket{00}_\logical{} &{} = (\ket{0000}+\ket{1111})/\sqrt{2}, \\
    \ket{01}_\logical{} &{} = (\ket{0011}+\ket{1100})/\sqrt{2}, \\
    \ket{10}_\logical{} &{} = (\ket{0101}+\ket{1010})/\sqrt{2},\\
    \ket{11}_\logical{} &{} = (\ket{1001}+\ket{0110})/\sqrt{2},
\end{array}
\end{equation}
\begin{equation}    
\begin{array}{r@{}l}
\label{eq:codwordsXX}
\ket{{+}{+}}_\logical{} &{} = (\ket{{+}{+}{+}{+}}+\ket{{-}{-}{-}{-}})/\sqrt{2}, \\
\ket{{+}{-}}_\logical{} &{} = (\ket{{+}{-}{+}{-}}+\ket{{-}{+}{-}{+}})/\sqrt{2}, \\
\ket{{-}{+}}_\logical{} &{} = (\ket{{+}{+}{-}{-}}+\ket{{-}{-}{+}{+}})/\sqrt{2}, \\
\ket{{-}{-}}_\logical{} &{} = (\ket{{-}{+}{+}{-}}+\ket{{+}{-}{-}{+}})/\sqrt{2},
\end{array}
\end{equation}
\begin{equation}    
\begin{array}{r@{}l}
\label{eq:codwordsZX}
\ket{{0}{+}}_\logical{} &{}= (\ket{0000}+ \ket{1100}+ \ket{0011}+ \ket{1111})/2,\\
\ket{{+}{0}}_\logical{} &{}= (\ket{0000}+ \ket{0101}+ \ket{1010}+ \ket{1111})/2, \\
\ketbell{}              &{}= (\ket{0000}+ \ket{0110}+ \ket{1001}+ \ket{1111})/2.
\end{array}
\end{equation}

\section{Error detection circuit in detail}\label{Supp_section:Detailed_QEC_circuit}

\begin{figure*}[h!]
    \centering
    \includegraphics[width = \textwidth]{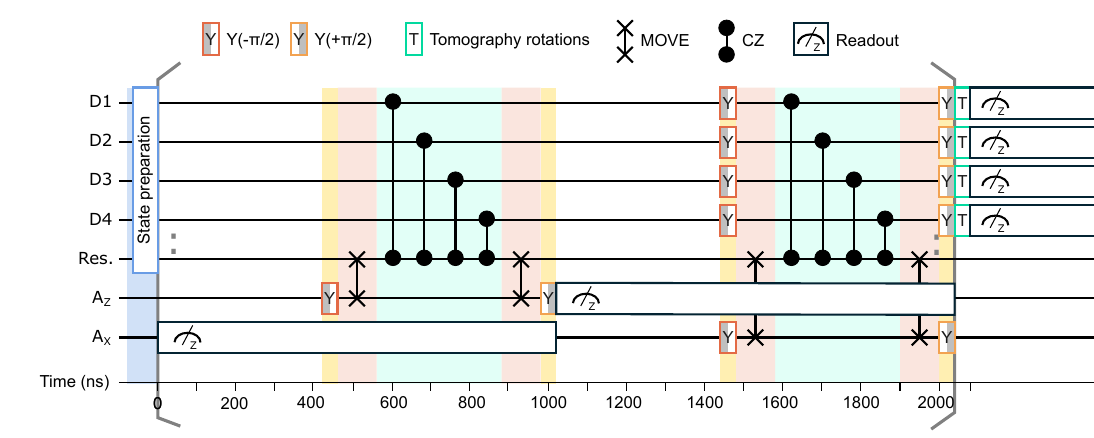}
    \caption{Representation of the error detection circuit shown in Fig.~\ref{fig:Sample_QEC_Cyle} with time on horizontal axis. The background colors serve as a guide to indicate the timing and duration of each operation.}
    \label{fig:Extended_cycle}
\end{figure*}

We show in Fig.~\ref{fig:Extended_cycle} error detection circuit, also shown in Fig.~\ref{fig:Sample_QEC_Cyle}, with the horizontal axis representing the time position of each operation.

\section{Logical shadow tomography}\label{Supp_section:RM}

Here, we further detail the tomography protocol that we use to extract logical and physical quantum properties of interest in our experiments.
The classical shadow framework has been widely used on different quantum platforms that are available today~\cite{Brydges2019, satzinger_realizing_2021, Yu2021, zhu_cross-platform_2022, Joshi2024observing, Vitale_QFI_2024, andersen2024thermalizationcriticalityanalogdigitalquantum}.
The state preparation of data qubits and $N$ cycles of error detection yield a state $\rho_\physical$.
We then apply unitary operations $U= \bigotimes_{i = 1}^4 U_i$ on the four data qubits where each $U_i$ is uniformly
sampled in $\left\{\mathbb{I}_2,\frac{1}{\sqrt{2}}\left(\begin{smallmatrix}
    1 & 1 \\
    1 & -1
\end{smallmatrix}\right),
\frac{1}{\sqrt{2}}
\left(\begin{smallmatrix}
    1 & -i \\
    1 & +i
\end{smallmatrix}\right) \right\}$, so that \mbox{$U_i^\dagger Z U_i=Z,X,Y$}, respectively (with $Z,X,Y$ being the Pauli matrices).
These rotations are Pauli basis rotations and form an unitary 2-design.

Then, we projectively measure the rotated state $U\rho_\physical U^\dag$ in the computational basis $\{\ket{\mathbf{d}}\}$, where $\ket{\mathbf{d}} = \ket{d_1, \dots, d_4}$ with $d_i \in \{0,1\}$. We repeat the randomized measurements $N_U N_M$ times, where $N_U$ is the number of sampled random unitaries $U^{(r)}$ and $N_M$ represents the number of measurements per unitary. The resulting dataset consists of $N_U N_M$ bitstrings, which we label as $\mathbf{d}^{(r,b)} = (d_1^{(r,b)}, \ldots, d_4^{(r,b)})$ for $r = 1, \dots, N_U$ and $b = 1, \dots, N_M$.

In addition to the data qubits, the experiments also involve projective measurements of the ancilla qubits, yielding stabilizer value $s^\xtype_n$ and $s^\ztype_n$.
Here, $n = 1, \dots, N$ labels the cycles of stabilizer measurements. 
In total, the dataset includes $2 N$ ancilla bitstring measurements for each applied unitary $U^{(r)}$.

We then perform a postselection on the stabilizer results which reduces the total number of data qubit bitstrings to $N_M'$ for each unitary $U^{(r)}$. The post-selected bitstrings are denoted as $\mathbf{d}^{(r,b')} = (d_1^{(r,b')}, \ldots, d_4^{(r,b')})$, with $r = 1, \dots, N_U$ and $b' = 1, \dots, N_M'$.

We construct, from this post-selected randomized measurement dataset, $N_U$ classical shadows~\cite{vermersch2023enhanced, Vitale_QFI_2024}
\begin{equation}
    \hat{\rho}^{(r)}_\physical
    =\sum_{\bf d} \hat{P}(\mathbf{d}|U^{(r)})\bigotimes_{j=1}^4 \left( 3\;{U^{(r)}_j}^{\dag} \ketbra{d_j}{d_j} U^{(r)}_j - \mathbb{I}_2\right) \label{eq:shadow}
 \end{equation}
where 
\begin{equation}
    \hat{P}(\mathbf{d}|U^{(r)}) = \sum_{b' = 1}^{N_M'} \frac{\delta_{\mathbf{d}, \mathbf{d}^{(r,b')}}}{N_M'} 
\end{equation}
is the estimated (noisy) Born probability from the randomized measurements dataset. This operator in Eq.~\eqref{eq:shadow} is an unbiased estimator of the underlying density matrix, i.e the average over the unitaries and the bitstring measurement results in $\mathbb{E}[\hat{\rho}^{(r)}_\physical] = \rho_\physical$~\cite{Huang2020}.
This property indeed allows us to perform quantum state tomography from the acquired dataset.

This operator defined in Eq.~\eqref{eq:shadow}, allows equally to construct \emph{logical shadows} for each applied random unitary $U^{(r)}$
\begin{equation}
    \hat{\rho}^{(r)}_\logical{} = 
    \sum_{i, j} \bra{i} \hat{\rho}^{(r)}_\physical \ket{j}/ P_\logical{} \label{eq:log_shadow}
\end{equation}
 with $\ket{i}$, $\ket{j}$ being the logical basis states defined in App.~\ref{Supp_section:CodeWord} and we define the estimator of the acceptance probability $P_\logical{} = \frac{1}{N_U}\sum_{r = 1}^{N_U} \Tr(\hat{\rho}^{(r)}_\physical)$.
 Based on the above property of the classical shadow, it follows straightforwardly that the average over the unitaries and measurement results provide an unbiased estimation of the underlying logical density matrix $\mathbb{E}[\hat{\rho}^{(r)}_\logical{}] = \rho_\logical{}$~\cite{Huang2020}.
 The operators defined in Eq.~\eqref{eq:shadow} and Eq.~\eqref{eq:log_shadow} are the building blocks for extracting properties of the prepared state.

We can define the shadow estimator of the fidelity \({F}\) of the quantum state with respect to pure state of interest $\ket{\psi_0}$ as~\cite{vermersch2023enhanced}
\begin{equation}
    {{ F}} = \frac{1}{N_U} \sum_{r = 1}^{N_U} \bra{\psi_0} \hat{\rho}^{(r)} \ket{\psi_0}, 
\end{equation}
where $\hat{\rho}^{(r)}$ can be either be the physical $\hat{\rho}^{(r)}_\physical$ or logical $\hat{\rho}^{(r)}_\logical{}$ shadow. 
Similarly, the unbiased estimator of the purity of the physical or logical quantum state is~Refs~\cite{Elben2020b, Rath_QFI_2021,vitale2022symmetry} 
\begin{equation}
    p_{2} = \frac{1}{N_U(N_U -1)} \sum_{r_1 \ne r_2} \Tr(\hat{\rho}^{(r_1)}\hat{\rho}^{(r_2)})  
\end{equation}
where $p_{2}$ can be either be the physical $p_{2,\physical{}}$ or logical $p_{2,\logical{}}$ purity.

\section{Density matrices}\label{Supp_section:logical_tomo}

Here, we provide the full density matrices from the tomography experiments. 

\begin{figure*}[h!]
    \includegraphics[scale=1.0]{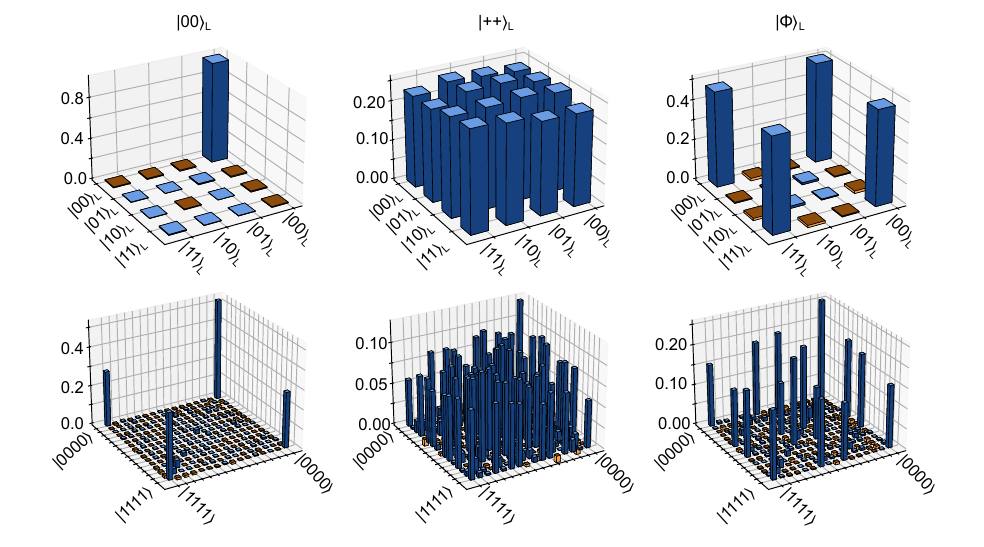}
    \put(-450,240){(a)}    
    \put(-310,240){(b)}
    \put(-160,240){(c)}    
    \put(-450,120){(d)}
    \put(-310,120){(e)}    
    \put(-160,120){(f)}
    \caption{Tomography for logical states $\ket{00}_\logical{}$ , $\ket{{+}{+}}_\logical$ and $\ketbell$. a-c) represents the tomography in the logical space of the two logical qubits and c-f) represents the tomography in of the physical space of the four data qubits for the respective mentioned states starting with the initial state of $\ket{0000}$, $\ket{{+}{+}{+}{+}}$ and $\ketbell$. Blue represents positive while yellow represents negative values of the real part of the density matrix.
    \label{fig:tomo}}
\end{figure*}

First, we show in Fig.~\ref{fig:tomo}a--c the density matrices of $\ket{00}_\logical{}$, $\ket{{+}{+}}_\logical$ and $\ketbell$ states obtained after 1 cycle of error detection. We expect 1, 16 and 4 density matrix elements with value 1, 0.25 and 0.5 for the 3 presented states correspondingly. Due to the high fidelity, other matrix elements are small and non-systematic. In Fig.~\ref{fig:tomo}d--f, we present density matrices of the physical state before projection to the logical subspace. Here we expect 4, 64 and 16 matrix elements with values 0.5, 0.125 and 0.25 correspondingly. In the case of the physical density matrices, we can clearly observe effects of qubit decay, which increases ground-state probability and reduces all other expected matrix elements. However, dephasing is not as visible, which reduces off-diagonal elements compared to diagonal elements.

\begin{figure*}[h!]
    \includegraphics[scale=1.0]{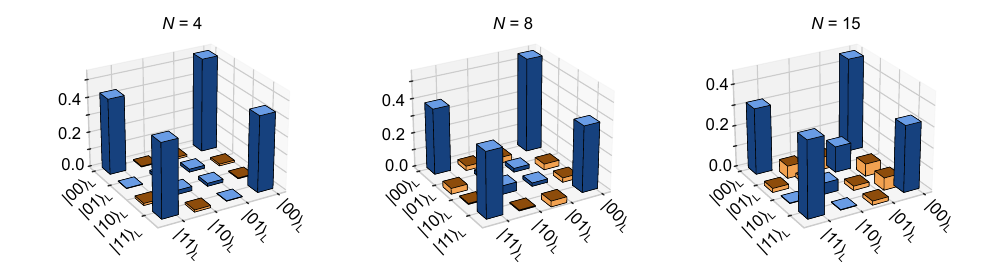}
    \put(-450,120){(a)}    
    \put(-300,120){(b)}
    \put(-140,120){(c)} 
    \caption{
    Tomography of the real part of the logical Bell density matrix for different number of cycles $N=4$ (a), $N=8$ (b) and $N=15$ (c).
    \label{fig:bell_tomo}}
\end{figure*}

For the Bell state we study the dynamics of the density matrix by presenting the density matrix after $N = 4, \, 8$ and $15$ cycles of error detection Fig.~\ref{fig:bell_tomo}. We observe an increase of unwanted diagonal elements indicates depolarizing noise. We also observe increased value of $\ket{00}\bra{00}_\logical$ independent of $N$ which is likely due to the qubit decay in the state preparation stage.

\pagebreak
\section{Syndrome detection probability}\label{Supp_section:detectors}

From the experiments described in Sections~\ref{sec:logical error rate} and~\ref{sec:bell} we already extracted the average probability of the first error happening in a given cycle $P_\mathrm{S}$. A complementary way to characterize the performances of quantum error detection is to study error detection probability, \textit{mean syndrome}, as a function of the error detection cycle number without prior postselection. 

For each error detection cycle $n$ we extract a bit $d_n$ for either stabilizers. For our circuit, where we do not reset ancilla qubits at the start of each cycle, the stabilizer value is $s_n = {+}1$ if $d_n \oplus d_{n-1}=0$ and ${-}1$ otherwise. The syndrome $\sigma(n) = (1-s_{n} \times s_{n-1} )/2$ is equal to $1$ if a change of parity has been detected between cycles $n-1$ and $n$, and $0$ otherwise.

\begin{figure*}[h!]
    \includegraphics[scale=1.0]{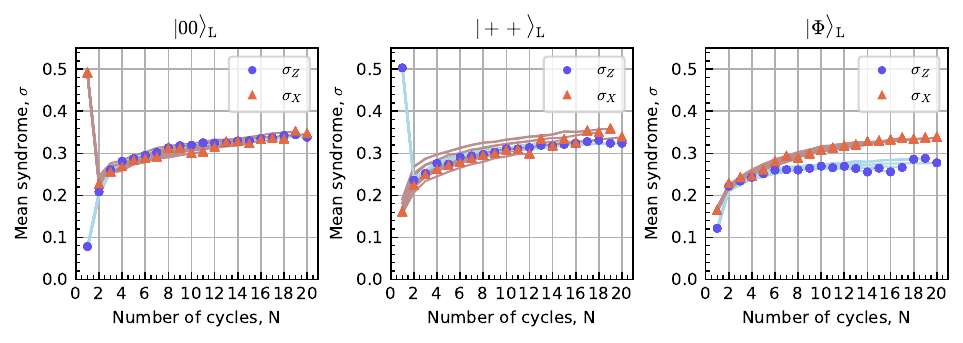}
    \put(-450,150){(a)}    
    \put(-300,150){(b)}
    \put(-140,150){(c)} 
    \caption{Mean syndrome $\sigma_{\xtype}(N,n)$ and $\sigma_{\ztype}(N,n)$ for $S_\xtype{}$ and $S_\ztype{}$ stabilizers repeated over cycles of error detection (with no postselection) for logical states $\ket{00}_\logical$ (a), $\ket{{+}{+}}_\logical$ (b) and $\ketbell$ (c). The data markers correspond to the measurement after the $N^{th}$ cycles and the nodes of the segmented lines correspond the mid-circuit measurements happening at every $n \in [1,N[$ cycles.
    The results are averaged over $10^5$ repetitions.}
    \label{fig:Syndromes}
\end{figure*}

We observe, that after the first cycle, the stabilizer that is not of the encoded basis triggers an error syndrome for about half of the experimental runs, see Fig.~\ref{fig:Syndromes}a,b, as expected from probabilistic encoding. Due to the deterministic encoding, both detection probabilities are low for the logical Bell state, see Fig.~\ref{fig:Syndromes}c.
After the first cycle, the mean syndrome reaches \num{0.3}, which is comparable to $1-P_{\mathrm{S}}$ observed in other experiments described in the main text. There is also a trend of increasing detection probability which we attribute this trend to leakage accumulation~\cite{Google2023overcomingLeakage}, see Section~\ref{Supp_section:Leakage}.

\section{Leakage detection and postselection}\label{Supp_section:Leakage}

\begin{figure*}[h!]
    \includegraphics[scale=1.0]{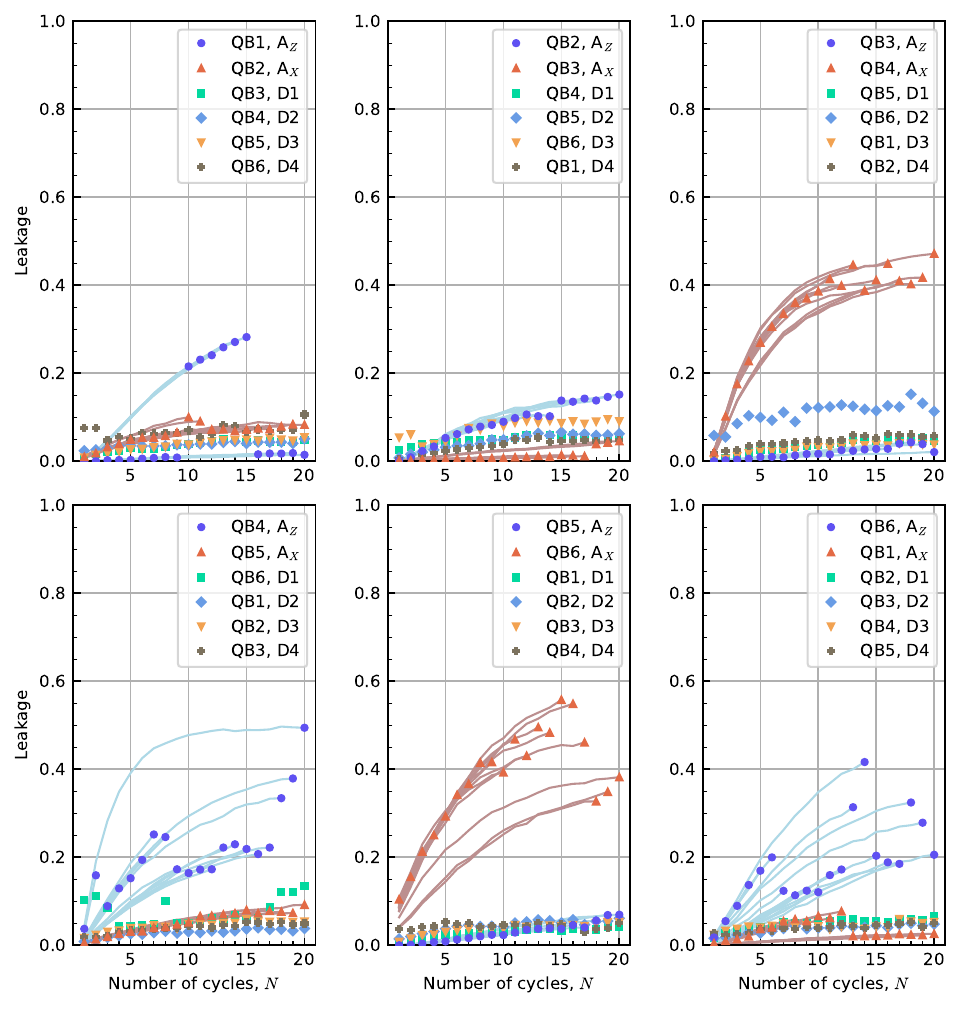}
    \put(-422,465){(a)}    
    \put(-270,465){(b)}
    \put(-118,465){(c)} 
    \put(-422,230){(d)}    
    \put(-270,230){(e)}
    \put(-118,230){(f)} 
    \caption{Measurement of leakage during repeated error detection cycles. The role of each physical qubit in the QED protocol is swapped from (a) to (f), as detailed in the legend. For ancilla qubits, the data markers correspond to the measurement after the $N^{th}$ cycles and the nodes of the segmented lines correspond the mid-circuit measurements happening at every $n \in [1,N[$ cycles. For data qubits, leakage is only measured after $N$ cycles. The data qubits are initally prepared in the $\ket{00}_\logical{}$ state.
    These results are averaged over $10^4$ repetitions.}
    \label{fig:Leakage}
\end{figure*}

To study possible reasons for detection probability dynamics, we measure the leakage accumulation in the qubits during the error detection cycles. While the rest of the experiments of this paper are done with a two state classifier, we use here a three state classifier and optimize the readout parameters to differentiate the three qubit states $\ket{0}$, $\ket{1}$ and $\ket{2}$. In this experiment, the correct assignment probability is $>99~\%$ for state $\ket{0}$ (except \data{}2 : $>\SI{97}{\percent}$), $>\SI{94}{\percent}$ for state $\ket{1}$, and $>\SI{93}{\percent}$ for state $\ket{2}$ (except $\ancilla{}_\xtype{}$ : $>84\%$). We account a qubit measured in the state $\ket{2}$ as leakage. 

We execute the quantum error detection protocol with the three state classifier.
In Fig.~\ref{fig:Leakage}, we show the leakage fraction detected for ancilla and data qubits. Moreover, we swap the role of physical qubits (see  Fig.~\ref{fig:Leakage} caption) to pin down any individual physical qubit effects.
In many cases, we see a steady increase of leakage fraction in ancilla qubits with the cycle index. In contrast, the leakage fraction in data qubits remains low and within the range of readout error.
In Fig.~\ref{fig:Leakage}(a), we observe an exceptionally high leakage for $\ancilla{}_\ztype{}$ between cycle \num{10} and cycle \num{15} that could be explained by some drift of the device parameters during the long measurement time of the experiment.

A probable source of leakage is state transition during qubit readout~\cite{Martinis2016Leakage,Blais2022Leakage}, since ancilla qubits are measured every cycle. Leakage induced by \move{} gates would also explain the results since they are only used between the ancilla qubits and the resonator. Further study of the physical properties of the qubits and the \move{} gates is expected to be beneficial to improve the performance of the QPU.

Leakage can be highly detrimental for quantum computation~\cite{varbanov2020leakage}. However, the fact that it seems contained in the ancilla qubits without propagating to the neighbors~\cite{Google2023overcomingLeakage} means we can mitigate its impact by resetting the ancilla qubits using one of the protocols that have already been demonstrated~\cite{Google2021Reset,Wallraff2018Reset,Filipp2018reset,Shengyu2021reset,Krinner2025LRU} without introducing additional errors into the data qubits.
Moreover, the [[4,2,2]] code can correct for one leakage error in a data qubit by treating it as an erasure error~\cite{Grassl_1997,AtomComputing2024}. Correcting for leakage error differs from passive leakage removal because correction makes use of the fact that we know when leakage happened. Correcting for leakage during fault tolerant memory requires a method to detect leakage without measuring the stored info in the qubit and this is not trivial with standard dispersive readout technique. Correcting for some leakage might be useful at the logical readout events, but by then it likely has caused stabilizer flips and errors on other data qubits. Alternatively, leakage can be transferred to the ancilla qubit and transformed into Pauli error~\cite{Google2021Reset,Google2024}.

Note that as we use a two state classifier in the rest of the experiments of this paper, leakage cannot be detected. As the readout parameters are optimized to best distinguish the states $\ket{0}$ and $\ket{1}$ in the IQ plane, the $\ket{2}$ states are mostly classified as $\ket{1}$. Therefore, leakage in ancilla qubits would likely be interpreted as an error in the stabilizer measurement and discarded by the postselection. For this reason, we believe that leakage does not affect the logical error rate even when a two state classifier is used. 

 \pagebreak
\section{Qubit specifications}\label{Supp_section:qubit_characteristics}

We show the main QPU characteristics in Table~\ref{table:QPUbenchmark}. For for more detailed description of the device, calibration and benchmarking methods, and definitions, refer to Ref.~\cite{Deneb}.

\begin{table*}[h!]
\centering
\begin{tabular}{llrrrrrrr}
\hline
Parameter  & Description & \qubit{}1 & \qubit{}2 & \qubit{}3  & \qubit{}4 & \qubit{}5 & \qubit{}6     & Res.    \\
\hline
$f_q$ (GHz)            &  Qubit/resonator frequency & $4.67$ & $4.47$ & $4.41$ & $4.52$ & $4.63$ & $4.93$ & $4.22$  \\
$T_\mathrm{q}$ (mK) &  Qubit temperature  & 46.3  & 42.0  & 43.6  & 40.9   & 43.0  & 45.8 &   \\
$\fidReadout$ &  Readout fidelity   & \SI{98.3 \pm 0.3}{}  & \SI{98.6 \pm 0.2}{}  & \SI{98.7 \pm 0.3}{}  & \SI{99.1 \pm 0.2}{}  & \SI{98.9 \pm 0.2}{}  & \SI{98.7 \pm 0.6}{} &   \\
$\fidSQBGind$ (\%)  &  Individual SQG fidelity  & \SI{99.93\pm0.02}{}  & \SI{99.94\pm0.04}{}  & \SI{99.96\pm0.02}{}  & \SI{99.96\pm0.01}{}  & \SI{99.96\pm0.01}{}  & \SI{99.89\pm0.3}{} &   \\
$\fidSQBGsim$ (\%)  &  Simultaneous SQG fidelity  & \SI{99.93\pm0.02}{}  & \SI{99.92\pm0.04}{}  & \SI{99.96\pm0.03}{}  & \SI{99.95\pm0.01}{}  & \SI{99.59\pm0.01}{}  & \SI{99.87\pm0.3}{} &   \\
$\fidMOVE$ (\%)  &  Double MOVE fidelity & \SI{99.11\pm0.05}{}  & \SI{99.34\pm0.03}{}  & \SI{99.00\pm0.03}{}  & \SI{99.30\pm0.03}{}  & \SI{98.31\pm0.06}{}  & \SI{97.95\pm0.10}{}   \\
$\fidCZ$ (\%)  &  CZ fidelity   & \SI{98.90\pm0.05}{}  & \SI{98.75\pm0.04}{}  & \SI{98.97\pm0.03}{}  & \SI{98.04\pm0.08}{}  & \SI{98.53\pm0.14}{}  & \SI{96.61\pm0.05}{}  \\
\hline
\end{tabular}
\caption{\label{table:QPUbenchmark} Characteristics of the QPU gates, component frequencies and temperatures, see~\cite{Deneb}.
}
\end{table*}

Because the QPU characteristics fluctuate in time, the experiments described in the main text were done in two different configurations that gave better performances at the time of the experiment. We carry out the calibration and benchmarking sequence twice a day for single qubit gates, two qubit gates and readout. One entire experiment, such as in Fig.~\ref{fig:Logical_expectation_values}, takes around 12h to run, which is the time between two recalibrations. Experiments described in Section~\ref{sec:logical error rate} use Configurations A, see Table~\ref{table:QPU parameters A}, while experiments presented in Sections~\ref{sec:stabilizer tomo}, \ref{sec:tomography} and \ref{sec:bell} were done in Configuration B, see Table~\ref{table:QPU parameters B}. We give the characteristic life time of each qubits measured in within the same day as the experiments. These values are used in the respective simulations of the experiment. 

\begin{table*}[h!]
\centering
\begin{tabular}{llrrrrrrr}
\hline
Parameter  & Description & \qubit{}1 & \qubit{}2 & \qubit{}3  & \qubit{}4 & \qubit{}5 & \qubit{}6     & Res.    \\
\hline
 &Role of the qubit  & $\ancilla{}_\ztype{}$ & $\ancilla{}_\xtype{}$ & \data{}1 & \data{}2 & \data{}3 & \data{}4 & Res.  \\
$T_1$ (µs) & Lifetime          & 26.1&	44.3&	64.5&	38.7&	40.8&	29.4& 5.4 \\
$T_2^* (µs)$ & Dephasing time & 45.1&	29.1&	34.7&	26.4&	47.2&	22.8& 10.3\\
$T_2^\mathrm{e}$ (µs) & Dephasing w.o. low freq noise contribution  & 51.3&	52.0&	45.3&	36.1&	56.0&	39.5& \\
\hline
\end{tabular}
\caption{\label{table:QPU parameters A} Characteristics of the QPU during experiments in Section~\ref{sec:logical error rate}. 
}
\end{table*}

\begin{table*}[h!]
\centering
\begin{tabular}{llrrrrrrr}
\hline
Parameter  & Description & \qubit{}1 & \qubit{}2 & \qubit{}3  & \qubit{}4 & \qubit{}5 & \qubit{}6     & Res.    \\
\hline
&Role of the qubit  & \data{}2 & \data{}3 & \data{}4 & $\ancilla{}_\ztype{}$ & $\ancilla{}_\xtype{}$ & \data{}1 & Res.  \\
$T_1$ (µs) & Lifetime          &25.7&	51.2&	59.1&	55.0&	49.0&	36.4& 5.7 \\
$T_2^* (µs)$ & Dephasing time & 37.9&	30.1&	31.5&	19.6&	60.3&	42.4& 11.9 \\
$T_2^\mathrm{e}$ (µs) & Dephasing w.o. low freq noise contribution & 42.4&	62.2&	43.0&	30.8&	60.9&	47.8&\\
\hline
\end{tabular}
\caption{\label{table:QPU parameters B} Characteristics of the QPU during experiments in Sections \ref{sec:stabilizer tomo}, \ref{sec:tomography} and \ref{sec:bell}.
}
\end{table*}

\section{Circuit simulation}\label{Supp_section:simulation}

We simulate the circuit using Stim version 1.14~\cite{gidney2021stim} to obtain the simulation results in Fig.~\ref{fig:Stabilizer_Characterization}, \ref{fig:Logical_expectation_values} and \ref{fig:Bell} of the main text. In the Stim circuit, the central resonator is modeled as a qubit and the \move{} gate as a SWAP gate.
We build an error model based on the error detection circuit and the performance of the QPU given in App.~\ref{Supp_section:qubit_characteristics}. 
For each gate we add a depolarization error detailed in Table~\ref{table:errors}.
We also add an idling error calculated from the time qubits stay idling $T_\mathrm{idl}$ and its $T_1$ and $T_2$ relaxation time~\cite{PhysRevA.86.062318}:

\begin{equation}
X_\mathrm{error} = Y_\mathrm{error} = \frac{1}{4} \left(1 - \exp(\frac{-T_\mathrm{idl}}{T_1})\right )
\text{,  }\quad
Z_\mathrm{error} = \frac{1}{2}\left(1 - \exp(\frac{-T_\mathrm{idl}}{T_2})\right ) - 
\frac{1}{4}\left(1
- \exp(\frac{-T_\mathrm{idl}}{T_1})\right ).
\label{eq:XYZerrors}
\end{equation}
A state preparation error {\ptherm} is added at the beginning of the circuit. It is calculated from the Boltzmann distribution $\ptherm = \exp(-\frac{\mathrm{hf_q}}{\mathrm{k_B\Ttherm}})$ using an effective temperature $\Ttherm$ estimated from single shot readout experiment, and frequency $f_q$ of each component.

\begin{table*}[h]
\setlength{\tabcolsep}{12pt}
\centering
\begin{tabular}{lll}
\hline
Operation    &   error probability & Stim operation name      \\
\hline

Unitary   & $\pSQBG = 2 (1-\fidSQBGsim)$                & \verb|DEPOLARIZE1|($\pSQBG$) \\
Hadamard  & $\pSQBG = 2 (1-\fidSQBGsim)$                & \verb|DEPOLARIZE1|($\pSQBG$) \\
\move{}   & $\pMOVE = \frac{4}{3}(1-\sqrt{\fidMOVE})$   & \verb|DEPOLARIZE2|($\pMOVE$) \\
CZ        & $\pCZ = \frac{4}{3}(1-\fidCZ)$              & \verb|DEPOLARIZE2|($\pCZ$)  \\
Readout   & $\pReadout = 1-\fidReadout$                 & \verb|M|($\pReadout$) \\
Idling    & Eq.~(\ref{eq:XYZerrors})                    & \verb|PAULI_CHANNEL_1|($X_\mathrm{error}$,$Y_\mathrm{error}$, $Z_\mathrm{error}$)  \\
Thermalization & $\ptherm = \exp(-hf/k_\mathrm{B}\Ttherm)$ & \verb|X_ERROR|(\ptherm) \\
\hline

\end{tabular}
\caption{\label{table:errors} List of error in Stim simulations and their associated error probability for each operation of the circuit. $h$ and $k_\mathrm{B}$ are the Planck and the Boltzmann constant respectively. We consider that $\fidMOVE$ is the fidelity of the double \move{} operation.
}
\end{table*}

\section{Error budget}\label{Supp_section:error_budget}

\begin{figure*}[h]
    \includegraphics[scale=1.0]{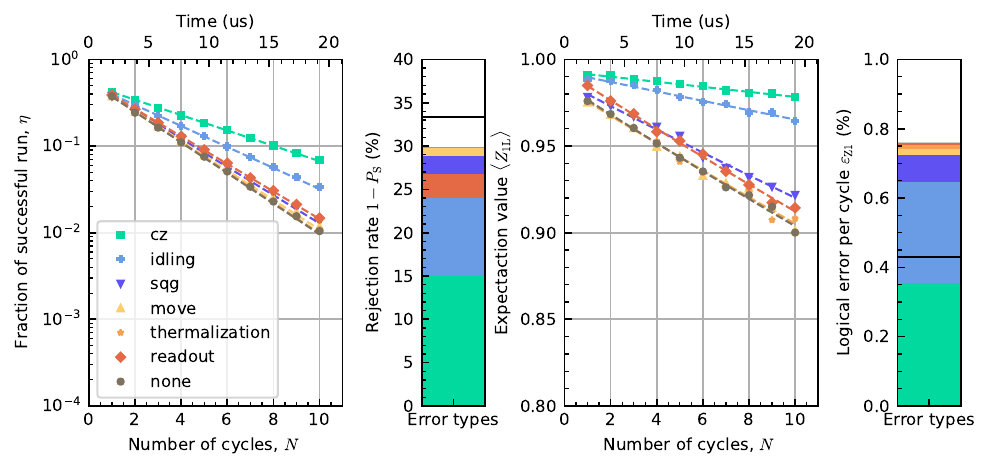}
    \put(-470,210){(a)}    
    \put(-300,210){(b)}
    \put(-230,210){(c)} 
    \put(-70,210){(d)} 
    \caption{Error budget made by simulating the error detection circuit by removing each time one error type indicated in the legend. "none" means that all the error types are considered.
    (a) Fraction of successful runs where no error have been detected. Dashed line: exponential fit to extract the rejection rate $1-P_\mathrm{S}$.
    (b) Individual contribution of each error type to the rejection rate. Black line rejection rate when all error types are considered.
    (c) Expectation value of the logical operator Z1 after the $N^{th}$ cycles. Dashed line: exponential fit to extract the logical error rate $\ler$ for the logical state.
    (d) Individual contribution of each error type to the logical error rate. Black line: logical error rate when all error types are considered.
    The simulation are calculated for the logical state $\ket{00}_\logical{}$ and logical operator Z1. The results are average over $10^5N$ simulation runs.}
    \label{fig:Error_budget}
\end{figure*}

To find the dominating error source, we simulate the error detection circuit by disabling one type of error each time.
We simulate the circuit for $N\in [1,10]$ cycle and fit the fraction of successful runs and the logical operator expectation value after the $N^{th}$ cycles to obtain the rejection rate $1-P_\mathrm{S}$ and logical error rate $\ler$, respectively, Fig.\ref{fig:Error_budget}. From these results, we calculate the individual contribution of each error type defined as the difference between the result where none are disabled and where this type is disabled.

We find that the logical error rate is dominated by the contribution of the CZ gate and idling errors. This is also reflected in their contribution to the rejection rate. The readout errors affects more the rejected rate than the logical error per cycle. False positive error detection events, due to readout errors, are much more likely to happen than false negative because false negatives can be caught by the next cycle while one false positive would reject the entire run.
We also notice a discrepancy between the simulated logical error rate where no error is disabled and the sum of each individual error contribution. This is due to the nonlinear dependency of the logical error rate on the physical error rate.

\section{Resonator errors and flag qubits}\label{Supp_section:Flag}

\begin{figure*}[h]
    \includegraphics[scale=1.0]{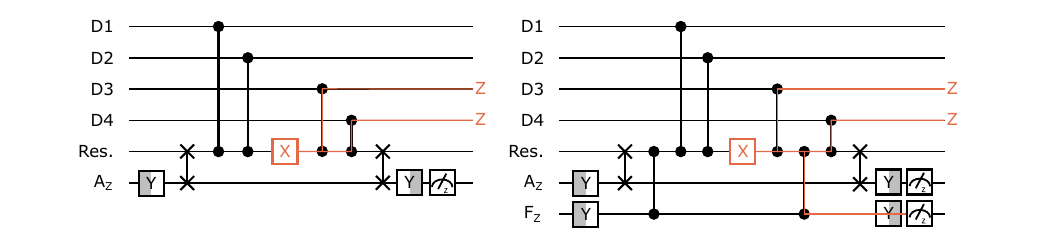} 
    \put(-480,100){(a)}    
    \put(-273,100){(b)}
    \caption{(a) Illustration of an X error in the resonator in the middle of a \ztype{} stabilizer circuit causing a Z error on two data qubits.
    (b) A variant of the same circuit incorporating a flag qubit F$_\mathrm{Z}$ enables the detection of the error.}
    \label{fig:Flag}
\end{figure*}

We investigate the consequence of errors in the resonator during the error detection cycle for the star architecture. In stabilizer codes, hook errors in the ancilla systems or stabilizer circuit can propagate to multiple data qubits and generate correlated errors~\cite{chao2018flag}. From the analysis of the circuit, we find that such hook errors can also be caused by bitflip errors in the resonator. For example, an X-error in the resonator, in the middle of the CZ sequence of the stabilizer circuit produces errors in multiple data qubits, specifically Z (X) errors if the resonator flip happens in the \xtype{} (\ztype{}) stabilizer circuit, see Fig.~\ref{fig:Flag}a.
The addition of a flag qubit~\cite{chao2018flag} alongside each ancilla qubit allows to detect these errors, see Fig.~\ref{fig:Flag}b.
In Appendix~\ref{Supp_section:Comparison}, we compare the performance of the error detection circuit with and without flag qubits.

\section{Simulation of higher weight stabilizers}\label{Supp_section:Comparison}

We use circuit level simulations to evaluate the error detection performance of star, flag-star and superdense stabilizer circuits with various weights. Superdense encoding circuit~\cite{gidney2023superdense} is an alternative method of simultaneously measuring \xtype{}- and \ztype{}-basis stabilizers supported by the same data qubits. We use Stim, see also~\ref{Supp_section:simulation}, to perform simulations of a circuit corresponding to a single stabilize pair and a single QEC cycle for weight- 4, 6, 8 and 10. The simulated star QPU has more data qubits and can, in princilpe encode $k$ logical qubits where $k+2$ is equal to the weight, effectively implementing the [[k+2,k,2]] code.
The circuit is adapted from Fig.~\ref{fig:Sample_QEC_Cyle}b and the flag-star circuit from Fig.~\ref{fig:Flag} by changing the number of CZ gate between the central resonator and the data qubits to match the weight.
Similarly, the superdense circuit is adapted from the weight-6 circuit of Ref.~\cite{Google2024ColorCode}.

To extrapolate the performance of a QPU containing more qubits than the one presented in this paper, we construct a uniform error model, see Table~\ref{table:Error_model}, corresponding to the average errors given in Tables~\ref{table:QPUbenchmark} and \ref{table:QPU parameters A}. The gate duration is as in the presented experiments described in  Fig.~\ref{fig:Extended_cycle}: 40~ns for single qubit gate, 80~ns for \move{}, 80~ns for \cz{} and 1~µs for readout.

\begin{table*}[h!]
\centering
\begin{tabular}{llrr}
\hline
Parameter                 & Description                     & \qubit        & Res.    \\
\hline
$f_q$ (GHz)             &  Qubit/resonator frequency        & $4.5$         & 4.2   \\
$T_\mathrm{q}$ (mK)     &  Qubit temperature                & 40            & 40    \\
$\fidReadout$  (\%)     &  Readout fidelity                 & 98.7          &       \\
$\fidSQBGsim$ (\%)      &  Simultaneous SQG fidelity        & 99.9          &       \\
$\fidMOVE$ (\%)         &  Double MOVE fidelity             & 98.8          &       \\
$\fidCZ$ (\%)           &  CZ fidelity                      & 98.3          &       \\
$T_1$ (µs)              & Lifetime                          & 40            & 5.4   \\
$T_2^*$ (µs)            & Dephasing time                    & 35            & 10    \\
$T_2^\mathrm{e}$ (µs)   & Dephasing w.o. low freq noise contribution & 40   & 15    \\
\hline
\end{tabular}
\caption{\label{table:Error_model} Uniform error model used in the simulations to extrapolate the performance of the architecture for higher-weight stabilizers. Each qubit characteristic is the average over every qubit of Table~\ref{table:QPUbenchmark} and Table~\ref{table:QPU parameters A}.}
\end{table*}

We run the simulation for $N\in [1,10]$ cycles for four data qubits prepared in the initial states $\ket{0000}$, $\ket{0011}$, $\ket{0101}$, $\ket{1001}$, $\ket{{+}{+}{+}{+}}$, $\ket{{+}{-}{+}{-}}$, $\ket{{+}{+}{-}{-}}$ and $\ket{{-}{+}{+}{-}}$. The additional data qubits are initialized in $\ket{0}$ for \ztype{} basis logical states or $\ket{+}$  for \xtype{} basis logical states.
We analyze the results of the simulations using the same error detection framework presented in Fig.~\ref{fig:Logical_expectation_values}. We postselect the measurement results based on the ancilla (and flag) qubit not detecting any errors and data qubits being assigned to the logical subspace. We fit the fraction of successful runs and the expectation values of the logical operators $Z_{\logical{}1}$, $Z_{\logical{}2}$, $X_{\logical{}1}$ and $X_{\logical{}2}$ to obtain the survival rate $P_\mathrm{S}$ and logical error per cycle $\ler_{\ztype{}1}$, $\ler_{\ztype{}2}$, $\ler_{\xtype{}1}$ and
$\ler_{\xtype{}2}$, respectively.

\begin{figure*}[h]
    \begin{overpic}[scale=1.0]{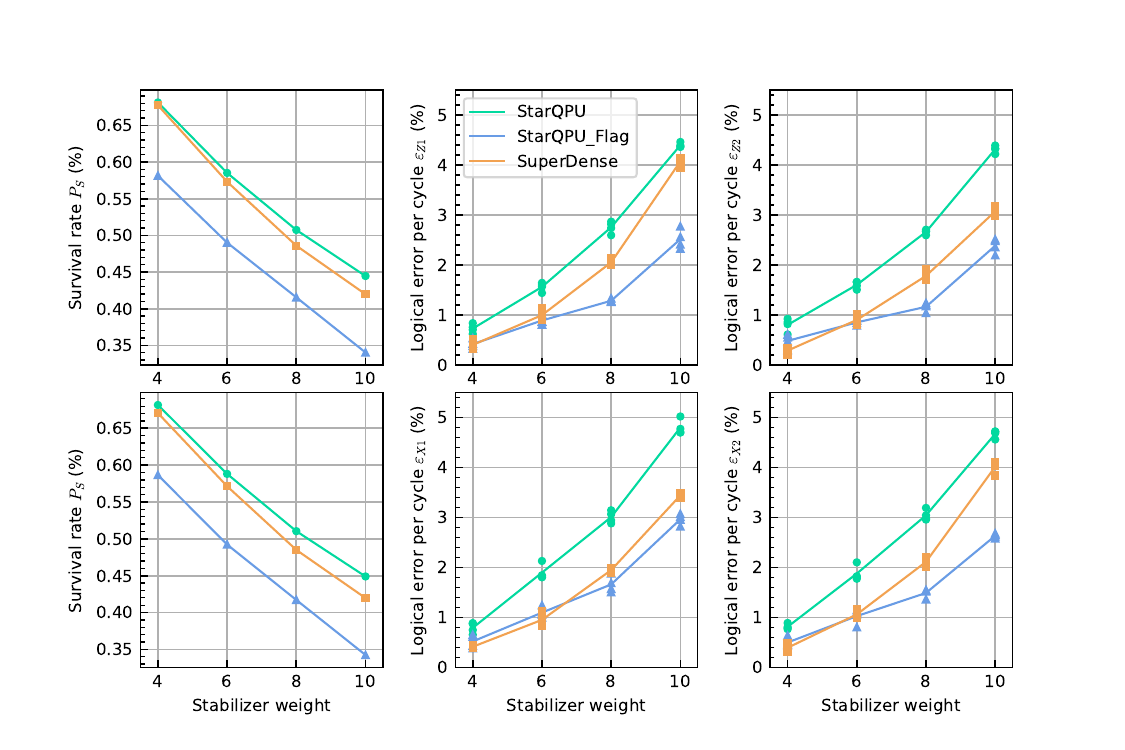} 
        \put(5,59){(a)}    
        \put(35,59){(b)}   
        \put(65,59){(c)}   
        \put(5,32){(d)}   
        \put(35,32){(e)}
        \put(65,32){(f)}
    \end{overpic}    
    \caption{Performance comparison of the error detection performances for the star QPU implementation with and without flag and for the superdense implementation. Post selection yield from two stabilizers (a,d), logical Z and X errors for various input states in the same basis for first (b,e) and second logical qubits (c,f).}
    \label{fig:Code_comparison}
\end{figure*}

The simulation results are shown in Fig.~\ref{fig:Code_comparison}. Unsurprisingly, $P_\mathrm{S}$ diminishes with increasing stabilizer weight since the number of error sources increases. Similarly, $\ler$ increases revealing higher probability of high-weight and undetected errors. We also notice that the flag-star accepts less runs than without flag. The difference corresponds mainly to the error detected in the resonator. Rejecting those error results in significantly smaller $\ler$, especially for higher stabilizer weights. This illustrates the significance of hook errors in the resonators and the importance of flag qubits to detect them.
Interestingly, while the superdense implementation always outperforms the star QPU without flag, the flagged implementation of the star QPU shows a better $\ler$ than the superdense implementation at high weight.
Note that the cycle time of both star circuits is at least twice the readout time (see Fig.~\ref{fig:Extended_cycle}). This leads to higher relative idling times at low weight in contrary to superdense implementation and explains the disadvantage of star-circuits for smaller stabilizer weights.

\begin{table*}[h!]
\centering
\begin{tabular}{lcccccccccccccccc}
\hline
Implementation (weight)           &  qubits   & SQG   &  \cz      & \move & cycle duration & $P_\mathrm{S}$ (\%)  &  $\ler_{Z0}$ (\%) & $\ler_{Z1}$ (\%)& $\ler_{X0}$ (\%)& $\ler_{X1}$ (\%)\\
\hline
Star QPU (4)            &   6       &   12  &   8       &   4   &  2.00 µs      & 0.68  & 0.71  & 0.80  & 0.79      & 0.81 \\
Star QPU (6)            &   8       &   16  &   10      &   4   &  2.00 µs      & 0.59  & 1.56  &  1.60  &  1.90    & 1.88 \\
Star QPU (8)            &   10      &   20  &   12      &   4   &  2.00 µs      & 0.51  & 2.75  &  2.66  & 3.01     & 3.04  \\
Star QPU (10)           &   12      &   24  &   20      &   4   &  2.04 µs      & 0.44  & 4.40  &  4.31  &  4.80    & 4.67  \\
\hline
Star QPU with flag (4)  &   8       &   16  &   12      &   4   &  2.00 µs      & 0.58  & 0.42  &  0.48  & 0.52     & 0.50 \\
Star QPU with flag (6)  &   10      &   20  &   16      &   4   &  2.00 µs      & 0.49  & 0.89  &  0.85  &  1.10    & 1.03 \\
Star QPU with flag (8)  &   12      &   24  &   20      &   4   &  2.04 µs      & 0.42  & 1.28  &  1.16  & 1.66     & 1.49 \\
Star QPU with flag (10) &   16      &   28  &   24      &   4   &  2.36 µs      & 0.34  & 2.52  & 2.38   &  2.96    & 2.63 \\
\hline
Superdense (4)          &   6       &   16  &   10      &   0   &  1.52 µs      & 0.68  & 0.40  &  0.29  & 0.41     & 0.39\\
Superdense (6)          &   8       &   20  &   14      &   0   &  1.68 µs      & 0.57  & 1.00  &  0.90  &  0.95    & 1.06\\
Superdense (8)          &  10       &   26  &   18      &   0   &  1.84 µs      & 0.49  & 2.05 &  1.78   & 1.95     & 2.11\\
Superdense (10)          & 13       &   30  &   22      &   0   &  2.00 µs      & 0.42  & 4.07  &  3.07  &  3.44   & 4.02 \\

\hline
\end{tabular}
\caption{\label{table:code_length}Number of qubits, gate count, cycle duration, $P_\mathrm{S}$ and $\ler$ for the different implementations (stabilizer weights) shown in Fig.~\ref{fig:Code_comparison}.}
\end{table*}

 \end{document}